\documentclass[twocolumn,amsmath,amssymb,10pt,superscriptaddress,a4paper,letterpaper,fleqn]{revtex4-1}
\usepackage{amssymb}
\usepackage{epsfig}
\usepackage{graphicx}
\usepackage{dcolumn}
\usepackage{array}
\usepackage{bm}
\usepackage{fancyheadings}
\usepackage{longtable}
\usepackage{multirow}
\usepackage{float}
\usepackage{afterpage}
\usepackage{color}

\setlongtables
\usepackage[breaklinks=true,linkbordercolor={1 1 1}]{hyperref}

\begin{document}
\setcounter{page}{1}


\title{
\qquad \\ \qquad \\ \qquad \\  \qquad \\  \qquad \\ \qquad \\
Accurate fission data for nuclear safety}

\author{A. Solders}
\affiliation{Physics and Astronomy, Applied Nuclear Physics division, Uppsala University, Box 516, SE-75120 Uppsala, Sweden}

\author{D. Gorelov}
\affiliation{Department of Physics, P.O. Box 35 (YFL), FI-40014 University of Jyväskylä, Finland}

\author{A. Jokinen}
\affiliation{Department of Physics, P.O. Box 35 (YFL), FI-40014 University of Jyväskylä, Finland}

\author{V.S. Kolhinen}
\affiliation{Department of Physics, P.O. Box 35 (YFL), FI-40014 University of Jyväskylä, Finland}

\author{M. Lantz}
\affiliation{Physics and Astronomy, Applied Nuclear Physics division, Uppsala University, Box 516, SE-75120 Uppsala, Sweden}

\author{A. Mattera}
\affiliation{Physics and Astronomy, Applied Nuclear Physics division, Uppsala University, Box 516, SE-75120 Uppsala, Sweden}

\author{H. Penttil\"a}
\affiliation{Department of Physics, P.O. Box 35 (YFL), FI-40014 University of Jyväskylä, Finland}

\author{S. Pomp}
\email[Corresponding author: ]{stephan.pomp@physics.uu.se}
\affiliation{Physics and Astronomy, Applied Nuclear Physics division, Uppsala University, Box 516, SE-75120 Uppsala, Sweden}

\author{V. Rakopoulos}
\affiliation{Physics and Astronomy, Applied Nuclear Physics division, Uppsala University, Box 516, SE-75120 Uppsala, Sweden}

\author{S. Rinta-Antila}
\affiliation{Department of Physics, P.O. Box 35 (YFL), FI-40014 University of Jyväskylä, Finland}

\date{\today} 

\begin{abstract}
{The Accurate fission data for nuclear safety (AlFONS) project aims at high precision measurements of fission yields, using the renewed IGISOL mass separator facility in combination with a new high current light ion cyclotron at the University of Jyv\"askyl\"a. The 30~MeV proton beam will be used to create fast and thermal neutron spectra for the study of neutron induced fission yields. Thanks to a series of mass separating elements, culminating with the JYFLTRAP Penning trap, it is possible to achieve a mass resolving power in the order of a few hundred thousands \cite{10Pent}.


In this paper we present the experimental setup and the design of a neutron converter target for IGISOL. The goal is to have a flexible design. For studies of exotic nuclei far from stability a high neutron flux ($10^{12}$ neutrons/s) at energies 1 - 30 MeV is desired while for reactor applications neutron spectra that resembles those of thermal and fast nuclear reactors are preferred. It is also desirable to be able to produce (semi-)monoenergetic neutrons for benchmarking and to study the energy dependence of fission yields. 

The scientific program is extensive and is planed to start in 2013 with a measurement of isomeric yield ratios of proton induced fission in uranium. This will be followed by studies of independent yields of thermal and fast neutron induced fission of various actinides.\\
}
\end{abstract}
\maketitle

\lhead{}
\chead{}
\rhead{}
\lfoot{}
\rfoot{}
\renewcommand{\footrulewidth}{0.4pt}


\section{ INTRODUCTION}

The fission product yield is an important characteristic of the fission process. In fundamental physics,
knowledge of the yield distributions is needed for a better understanding of the fission process itself. For nuclear energy applications good knowledge of neutron induced fission product yields is crucial in many aspects, including criticality and reactivity calculations for reactor design, dosimetry and fission gas production for reactor safety and improved burn-up predictions \cite{IAEA-TECDOC-1168}. For the management of the nuclear waste, i.e. depositories, reprocessing, transmutation and so on, as well as for Generation IV reactor concepts, good knowledge of the composition of the spent fuels is required. The inventory of fission products in a reactor determine the decay heat, both residual heat after reactor shut down and the decay heat from spent fuel. Typically, the decay heat is dominated by fission products for the first 50-80 years after extraction of spent nuclear fuel from a reactor \cite{05Milles}.

The successful operation of nuclear power plants shows that the current knowledge of the underlying nuclear physics processes is generally sufficient. Predictions of macroscopic reactor parameters with model codes, as well as calculations of the isotopic composition of spent nuclear fuel, are in reasonable agreement with reality. Nevertheless, more accurate nuclear data would improve the predictions of fuel compositions and hence both safety and fuel economy.

In a supplement to WRENDA 93/94 - World Request List for Nuclear Data \cite{IAEA-INDC(SEC)-104} the International Atomic Energy Agency (IAEA) concludes that for independent neutron induced fission yields practically all fissioning systems need to be further investigated and that it is recommended to measure the energy dependence of yields for neutron energies ranging from thermal to very high \cite{IAEA-INDC(SEC)-105}. Other measurements are needed to improve model calculations of for example the even-odd effect of the fissioning nuclide. Independent fission yield measurements near symmetry are also needed to improve semi-empirical model parameters.

\section{The experimental technique}

The AlFONS project will use the upgraded IGISOL-JYFLTRAP facility at the accelerator laboratory of the University of Jyv\"askyl\"a. With the Ion Guide Isotope Separator On-Line (IGISOL) technique high yields of fission products are selected and then mass separated in the Penning trap JYFLTRAP. This method has proven to be very useful for the determination of independent fission yields. So far experiments have been performed with 25 MeV protons on $^{232}$Th and $^{238}$U, with 50 MeV protons on $^{238}$U and with 25 MeV deuterons on $^{238}$U. These measurements are described in more detail elsewhere in these proceedings \cite{13Pent}.

\subsection{The IGISOL-JYFLTRAP facility}
The recent move of the IGISOL-JYFLTRAP facility to a new experimental area includes a general upgrade. An important addition is the new MCC30/15 cyclotron \cite{mcc30} that can provide protons in the energy range 18 - 30 MeV, and deuterons of 9 - 15 MeV. A beam current of up to 100 $\mu$A is expected, making it possible to consider high intensity neutron beams. In combination with the heavy ion K-130 cyclotron, up to 4000 beam hours per year could be delivered to IGISOL.

\begin{figure}[!htb]
\includegraphics[width=0.80\columnwidth]{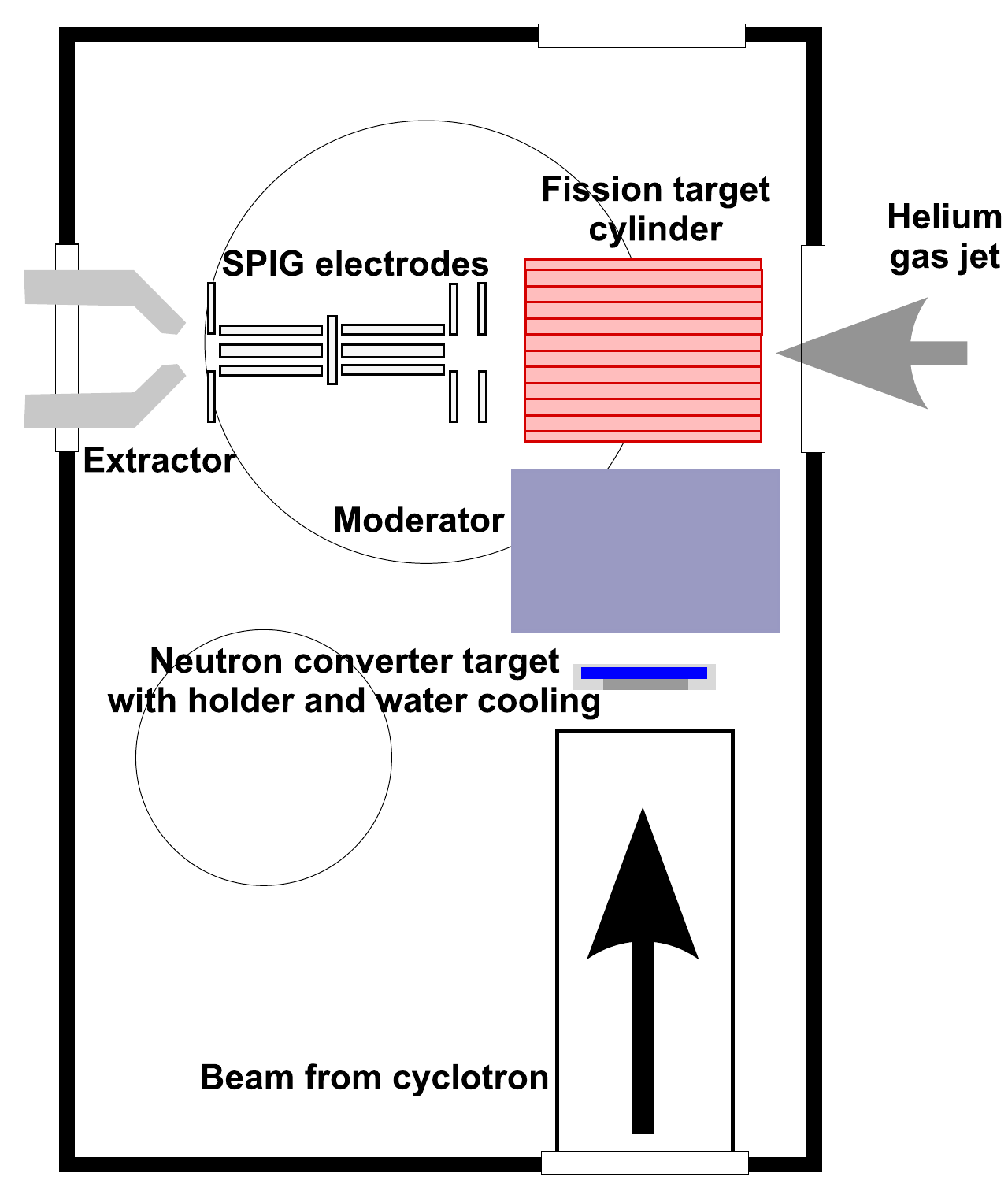}
\caption{Sketch of the reaction chamber with the proposed setup for neutron induced fission yield measurements. Not to scale.}
\label{fig3}
\end{figure}

In the reaction chamber (Fig.~\ref{fig3}) the beam from the cyclotron is impinged on a fissionable target. The ionized fission products leave the thin target and are retarded in a stream of noble gas. In this stream the ions are transported to a sextupole ion guide (SPIG) and accelerated for further transport down the beam line. The neutral gas is not affected by the electric field of the SPIG and is effectively pumped away between the SPIG rods. Thanks to the high ionization potential of the buffer gas, usually Helium, a large fraction of the ions are extracted from the reaction chamber as singly charged \cite{10Pent}.

The ions are accelerated to 30 keV and isobarically separated in a 55$^{\circ}$ dipole magnet with a mass resolving power of 500. In the RFQ buncher and cooler \cite{02Niem} the ions are accumulated and their energy spread is reduced, using buffer gas cooling, before they are further transported to the Penning trap system JYFLTRAP. In the purification part of JYFLTRAP the mass separated fission products are further cooled and the magnetron orbits are expanded by applying a mass independent dipole excitation. In the final step a mass selective quadrupole excitation is used to re-center those fragments which cyclotron frequency, $f_c=\frac{1}{2\pi }\frac{qB}{m}$, match the excitation frequency. These are then extracted from the trap through a 2~mm aperture and detected with an MCP detector. With this scheme a mass resolving power of more than $10^{5}$ has been demonstrated \cite{10Pent}. For a detailed description of the purification process see for example reference \cite{04Kolh}.
 
\subsection{The neutron converter target}
\begin{figure*}[!htb]
\includegraphics[width=0.95\textwidth]{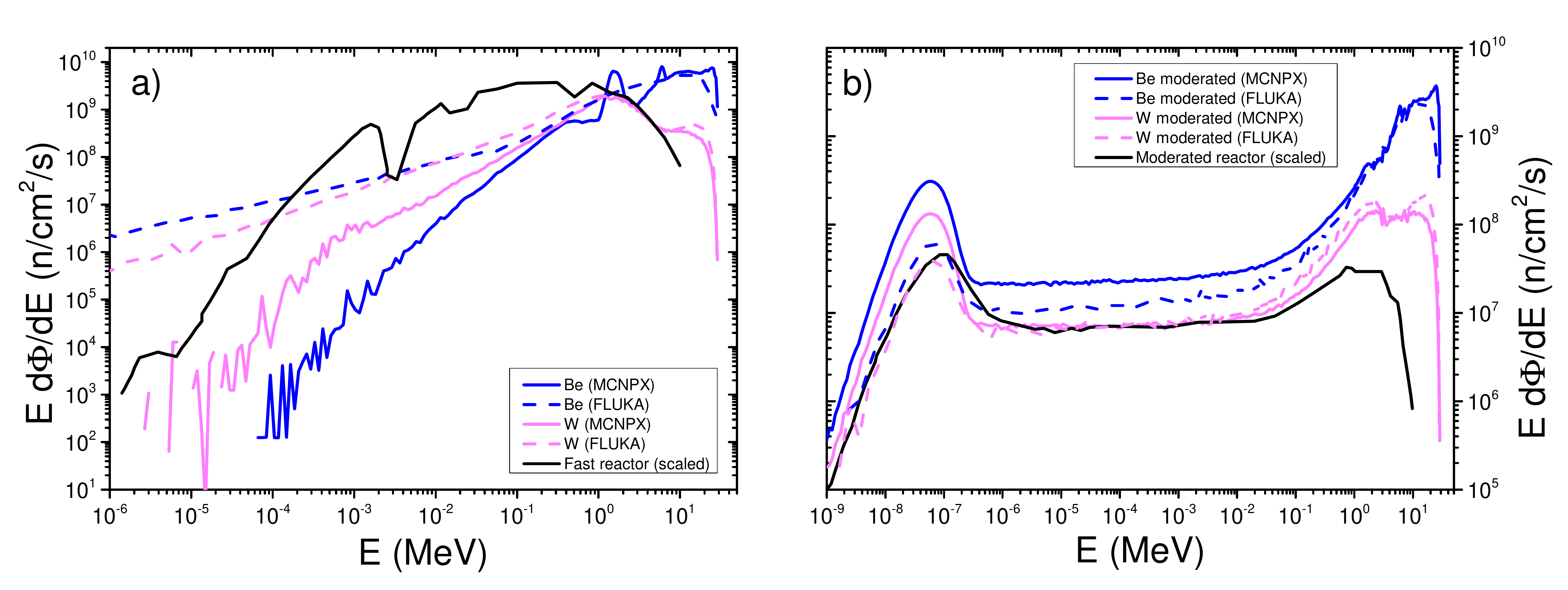}
\caption{Simulated spectra from beryllium och tungsten using FLUKA and MCNPX; a) shows the obtained fast spectra together with a typical fast reactor spectrum \cite{USDE} while b) shows the moderated spectra and a typical LWR spectrum \cite{USDE}. The reactor spectra has been arbitrary rescaled for comparison.}
\label{fig4}
\end{figure*}

So far only proton induced independent fission yields have been investigated with the Penning trap technique. For the  study of neutron induced fission a neutron converter target is being developed. As a first approach a design similar to the ANITA target \cite{09Prok} at The Svedberg Laboratory (TSL) was considered. This is a thick tungsten target producing a white neutron spectrum through the W(p,xn) reaction. To be competitive compared to other neutron facilities in the studies of nuclides far from the line of stability, the converter should deliver $10^{12}$ high energy neutrons (above 1~MeV) per second on the fission target. For this reason also beryllium was considered as target material. To estimate the neutron flux for different target options the Monte Carlo codes FLUKA \cite{07Batt, 05Ferr} and MCNPX \cite{06Mcki} have been used. In Fig.~\ref{fig4} the obtained neutron spectra for 5\ mm targets of tungsten and beryllium is plotted. From this beryllium was selected  as the main option because of the higher neutron yield in the high energy regime.

For the studies of independent fission yields for nuclear power applications the neutron energy spectrum should preferably resemble that of nuclear reactors, either moderated light water reactors (LWR) or fast reactors. Alternatively, a close to mono-energetic spectrum could be used. In Fig.~\ref{fig4} simulated spectra from tungsten and beryllium with and without moderator is compared to typical spectra from LWR and fast breeder reactors. The conclusion from these simulations is to have a flexible design that can be adopted to different materials and thicknesses.



In order to benchmark the simulated neutron spectra and to try to resolve the discrepancies between the two Monte Carlo codes (Fig.~\ref{fig4}) a measurement of the neutron spectrum has been performed at TSL. For these measurements a prototype assembly, similar to that of the LENS target \cite{08Lave} at the Indiana University Cyclotron Facility (IUCF), has been constructed. The prototype holds a 5~mm beryllium target in front of a 10~mm thick layer of cooling water and an aluminum back plate. The target thickness is chosen so that the protons do not fully stop in the target but instead terminate in the cooling water. This drastically reduces the cooling requirements and avoid hydrogen buildup in the target. The trade-off is a small reduction of the neutron yield of about 5\%. More information about the neutron converter target and the measurement at TSL can be found elsewhere in these proceedings \cite{13Matt}.

\section{Scientific program}

\subsection{Isomeric yield ratios}
Many isomers exist among fission products and play an important role for the calculation of the decay heat after reactor shutdown. Furthermore, the beta delayed neutron emission probability from the isomeric state can be an order of magnitude different from that of the ground state. Thus, proper simulation of the effect of delayed neutrons in reactors requires accurate knowledge of the population of isomeric states in fission. Measurements of isomeric yield ratios are also important for simulations of the astrophysical r-process.

Up till now the most common method of determining yield ratios has involved radiochemical separation of fission products and subsequent radioactivity counting. With this method the possible isomeric pairs were limited to those that were located close to stability and shielded by stable or long-lived isotopes from production via the beta decay of more neutron rich isotopes. With the IGISOL-JYFLTRAP facility direct determination of the isomeric fission yield ratios by means of counting the fission product ions is possible. It has been demonstrated that the mass resolving power is sufficient to, for example, resolve the two isomers of $^{96}$Y separated by 1.4 MeV \cite{10Pent}.

It is also possible to further increase the resolution by the use of a dipole excitation in the precision trap. With this method a mass resolving power of $8\times 10^5$ has been achieved in the separation of $^{133m}$Xe from its ground state, only 233~keV below \cite{10Pera}. It is anticipated that one of the first measurements with the renewed IGISOL-JYFLTRAP facility, once in operation later this year, will be a measurement of a number of isomeric yield ratios of 25~MeV proton induced fission in $^{238}$U.

\subsection{Neutron induced fission product yields}
\begin{figure}[!htb]
\includegraphics[width=0.95\columnwidth]{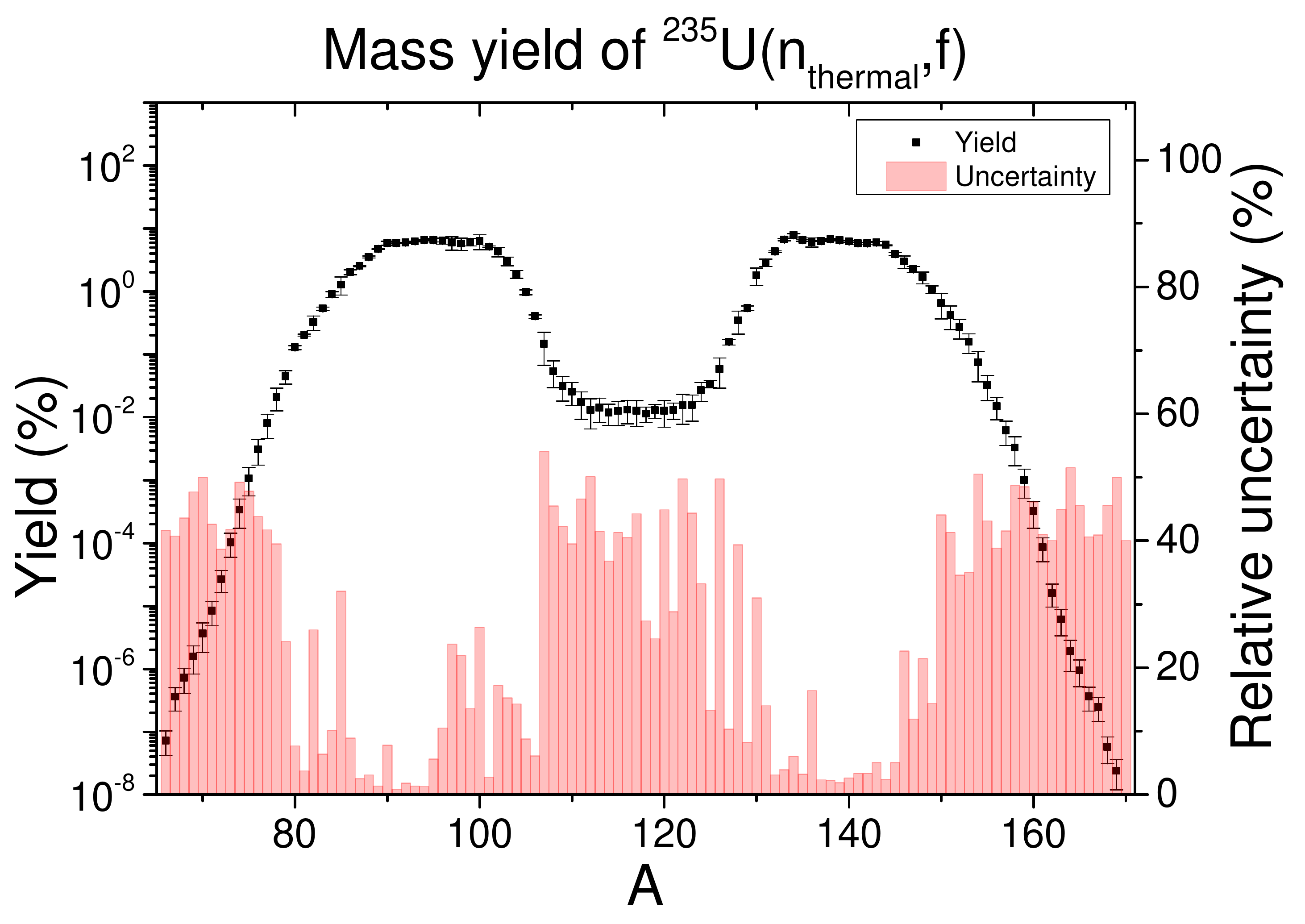}
\caption{Mass yield distribution of $^{235}$U neutron induced fission at thermal energy (black squares) from the ENDF/B-VII.1 library \cite{endf} together with relative uncertainties of the thermal yields from the same library (bars).}
\label{fig1}
\end{figure}

The well known shape of the fission yield mass distribution (Fig.~\ref{fig1}) with peaks around mass number 95 and 135 is characteristic for the thermal neutron induced fission of $^{235}$U in a light water reactor (LWR). This is the most investigated yield set (combination of fissile nucleus and neutron energy) and the mass yields in the peak regions are known within a few percent. However, in the wings and the valley the yields are generally not known to better than 40\% (relative uncertainty). In the latest ENDF/B-VII.1 evaluation of the thermal neutron fission of $^{235}$U only 106 individual yield ratios, out of the 998 tabulated, have relative uncertainties smaller than 10\% \cite{endf}.

For other yield sets the situation is generally worse. In the case of thermal fission of $^{233}$U, important for the Thorium fuel cycle, only 45 fission products are tabulated with uncertainties below 10\% while for fast neutrons no independent yields are presented at this precision \cite{endf}. 


Fission yields vary with incident neutron energy and for the development of a model to predict the energy dependence of fission yields systematic measurements at different energies are required. The measurements that exists are too scarce for reliable predictions and in many cases yields have only be measured at one or a few energies. There are, for example, no independent yield measurements of neutron induced fission in $^{241}$Pu above 5~MeV, $^{242m}$Am above thermal energies and $^{243}$Am at thermal energies.


The aim of the AlFONS program is to provide reliable measurements of independent fission yields of various actinides relevant for present and future reactor concepts. This includes improving the measurements of already studied yield sets as well as expanding the measurements to energies and actinides that have not yet been measured. 

It has been estimated that fission products with independent yields of $10^{-2}$ can be obtained at a rate of about 100 counts/s. In this case, an accuracy of 3\% (relative statistical uncertainty) can be obtained within 5 minutes. Lower yields need longer measurement times and in the best case yields of $10^{-5}$ with a 30\% statistical uncertainty can be achieved within about one hour measurement time.

Once the neutron converter has been commissioned the plan is to start with neutron induced fission of $^{238}$U using the 5~mm water cooled beryllium target. From there a number of parameters can be varied such as converter material, converter thickness and fission target material. It is also possible to vary the cyclotron parameters, energy and particle type (p or d) and with suitable moderator material we hope to obtain neutron spectra that resembles those of nuclear reactors. It is also possible to use thin converter targets to provide semi-monoenergetic neutron beams for reference measurements.

As stated in the introduction, almost all yield sets need further investigation and the plan is to start with easily obtainable actinides at well known energies to benchmark the method. The plan is then to expand to other energies and targets, limited only by availability and time. 



This work was supported by the European Commission within the Seventh Framework Programme through Fission-2010-ERINDA (project no. 269499), by the Swedish Radiation Safety Authority (SSM), by the Swedish Nuclear Fuel and Waste Management Co. (SKB), and by the Academy of Finland under project No. 139382 and the Finnish Centre of Excellence Programme 2012-2017 (Nuclear and Accelerator Based Physics Research at JYFL).


\end{document}